\documentclass[pdflatex,sn-mathphys-num]{sn-jnl}

\usepackage{multirow}   
\usepackage{graphicx}   
\usepackage{xcolor}     
\usepackage{float}
\usepackage{placeins}
\usepackage{tabularx}
\usepackage[labelfont=bf]{caption} 
\usepackage{amsmath,amssymb,amsfonts}%
\usepackage{amsthm}%
\usepackage{mathrsfs}%
\usepackage[title]{appendix}%
\usepackage{xcolor}%
\usepackage{textcomp}%
\usepackage{manyfoot}%
\usepackage{booktabs}%
\usepackage{algorithm}%
\usepackage{algorithmicx}%
\usepackage{algpseudocode}%
\usepackage{listings}%
\usepackage{anyfontsize}
\usepackage{braket}
\usepackage{subcaption}
\usepackage{longtable}
\usepackage{cleveref}
\usepackage{tikz}
\usepackage{booktabs}%
\usepackage{array}
\usepackage{quantikz}
\usetikzlibrary{arrows.meta, positioning, calc}

\tikzstyle{block} = [rectangle, rounded corners, minimum width=3.8cm, minimum height=1.2cm, text centered, draw=black]
\tikzstyle{classic} = [block, fill=gray!15]
\tikzstyle{quantum} = [block, fill=blue!10]
\tikzstyle{arrow} = [thick,->,>=stealth]

\begin{document}
\title{A Quantum Bagging Algorithm with Unsupervised Base Learners for Label Corrupted Datasets}


\author*[1]{\fnm{Neeshu} \sur{Rathi}}\email{neeshu\_r@ma.iitr.ac.in}

\author[1,2]{\fnm{Sanjeev} \sur{Kumar}}\email{sanjeev.kumar@ma.iitr.ac.in}
\equalcont{These authors contributed equally to this work.}


\affil*[1]{\orgdiv{Department of Mathematics}, \orgname{Indian Institute of Technology Roorkee}, \orgaddress{ \city{Haridwar}, \postcode{247667}, \state{Uttrakhand}, \country{India}}}

\affil[2]{\orgdiv{Mehta Family School of Data Science and Artificial Intelligence}, \orgname{Indian Institute of Technology Roorkee}, \orgaddress{ \city{Haridwar}, \postcode{247667}, \state{Uttrakhand}, \country{India}}}



\abstract{The development of noise-resilient quantum machine learning (QML) algorithms is critical in the noisy intermediate-scale quantum (NISQ) era. In this work, we propose a quantum bagging framework that uses QMeans clustering as the base learner to reduce prediction variance and enhance robustness to label noise. Unlike bagging frameworks built on supervised learners, our method leverages the unsupervised nature of QMeans, combined with quantum bootstrapping via QRAM-based sampling and bagging aggregation through majority voting. Through extensive simulations on both noisy classification and regression tasks, we demonstrate that the proposed quantum bagging algorithm performs comparably to its classical counterpart using KMeans while exhibiting greater resilience to label corruption than supervised bagging methods. This highlights the potential of unsupervised quantum bagging in learning from unreliable data.}




\maketitle
\section{Introduction}
In recent years, machine learning (ML) techniques have been increasingly leveraged to investigate some of the more challenging problems in quantum information science. Applications include tasks such as quantum state identification~\cite{lu2018separability,harney2020entanglement,ahmed2021classification}, reconstruction of quantum states~\cite{ahmed2021quantum}, precise estimation of parameters~\cite{wang2019machine}, and a variety of other inference and optimization scenarios~\cite{niu2019universal,zhang2019does,porotti2019coherent,bukov2018reinforcement,ding2021breaking,cirstoiu2020variational,schuff2020improving,lohani2022data}. The motivation for applying ML in these contexts stems from its ability to approximate solutions where classical numerical methods become inefficient—particularly in the presence of non-convexity, high-dimensional constraints, or limited computational resources.\\
Quantum Machine Learning (QML) is an emerging area of research that explores how quantum algorithms can be used to solve machine learning tasks more efficiently than their classical counterparts~\cite{biamonte2017quantum}. In principle, access to a large number of fully entangled qubits would allow QML algorithms to demonstrate a computational advantage over classical supercomputers across various application domains. However, current quantum devices operate within the constraints of the Noisy Intermediate-Scale Quantum (NISQ) era~\cite{preskill2018quantum}, where hardware is characterized by limited qubit counts and significant susceptibility to noise.
On gate-based quantum platforms, the reliability of computations deteriorates as circuits become deeper and more complex, making it difficult to encode large datasets without introducing substantial error. To ensure feasibility on today’s hardware, quantum circuits must be designed with reduced depth and gate complexity. Despite these limitations, careful design of shallow quantum models opens a path toward solving meaningful problems in practice, serving as a stepping stone toward realizing the broader potential of quantum technologies.\\
\noindent
Ensemble learning has become a fundamental strategy in classical machine learning to address the limitations of individual learners. By aggregating the predictions of multiple models, ensemble methods can reduce variance, improve robustness, and achieve better generalization performance. Among the most widely adopted ensemble techniques are Bagging~\cite{breiman1996bagging} and Boosting~\cite{freund1997boosting}, which respectively focus on reducing variance and bias by training base learners on different data distributions or by sequentially correcting prior errors.

\noindent
As quantum machine learning (QML) continues to develop, several quantum algorithms—such as quantum principal component analysis~\cite{qpca}, quantum support vector machines~\cite{qsvm}, and quantum $k$-nearest neighbors~\cite{qknn}—have demonstrated promising theoretical capabilities. However, real-world deployment of QML remains limited by hardware noise, qubit decoherence, and circuit depth restrictions. These limitations often result in unstable quantum models whose predictions vary significantly under small changes, motivating the need for ensemble techniques adapted to the quantum setting.
The study of quantum ensemble learning is still evolving, with several approaches focusing on bagging-, boosting-, and stacking-inspired strategies in both hybrid and fully quantum models. 

\noindent
Quantum ensemble methods have been also developed. For instance, Xie et al.~\cite{xie2017quantum} introduced a quantum-inspired bagging method called QI-Forest, which used randomized subsets and ensemble decision trees for regression tasks on UCI datasets. While their method showed improved prediction accuracy, it did not involve quantum processing and relied entirely on classical postprocessing. 
Khadiev et al.~\cite{khadiev2021quantum} focused on hybrid quantum-classical bagging using amplitude amplification to enhance classical decision trees. Their framework was conceptually sound but lacked empirical benchmarking and did not include circuit-level demonstrations, leaving scalability and robustness untested.
Incudini et al.~\cite{incudini2023resource} conducted empirical studies on both quantum bagging and boosting with Quantum Neural Networks (QNNs). They evaluated their methods on multiple datasets and reported gains in regression and classification tasks. However, their experiments used only a limited number of QNN variants. \\
While Schuld and Petruccione\cite{schuld2018quantum}, and Abbas et al.~\cite{abbas2020quantum} introduced a theoretical quantum ensemble framework based on Bayesian Model Averaging (BMA), where predictions are aggregated from a large set of non-trainable quantum classifiers. Their approach is based on the assumption that a sufficiently diverse ensemble of weak learners can collectively yield strong predictive performance. However, their work did not define concrete weak learners or offer experimental results, making it difficult to assess practical benefits or performance under noise. Instead, Araujo and da Silva~\cite{araujo2020quantum} proposed a quantum ensemble composed of trainable models. Their approach specifically utilizes a coherent superposition of quantum classifiers, where each base learner is implemented as a quantum neural network.
Macaluso et al.~\cite{macaluso2024efficient} proposed a fully quantum bagging approach that is based on bagging and is characterized by an exponential growth of the ensemble size at the price of a linear increase in the circuit depth. While the method reports perfect accuracy on some binary classification pairs, its performance on more challenging class distinctions (e.g., Iris 1 vs 2) remains modest, and the evaluation lacks testing on regression problems, leaving its applicability to continuous prediction tasks unexamined.\\
Eventually,
Yu et al.~\cite{yu2024quantum} proposed a hybrid ensemble framework combining quantum support vector machines (QSVMs) with classical bagging. Their experiments covered multiple binary classification datasets and showed stable performance. Nonetheless, their model remained partially classical and lacked a quantum subsampling procedure, which limited quantum coherence across learners.
Srikumar et al.~\cite{srikumar2024kernel} extended decision tree ensembles with quantum-enhanced kernel evaluations. Their method improved decision boundaries using quantum kernel estimation but did not include bootstrapping or fully quantum ensemble integration.
Tolotti et al.~\cite{tolotti2024ensembles} presented a comprehensive benchmark of quantum ensemble techniques—bagging, boosting, and stacking—using various base learners such as quantum cosine classifiers and quantum distance classifiers. Their hybrid models achieved competitive performance on 11 datasets. However, their focus remained on benchmarking, and the ensemble strategies did not incorporate quantum memory or state preparation mechanisms.
Yalovetzky et al.~\cite{yalovetzky2024qc} proposed a hybrid learning architecture combining supervised quantum $k$-means clustering with decision trees. They evaluated their model on both classification and regression datasets and showed that the hybrid approach improves training time. Yet, their architecture still relied on classical preprocessing and lacked a fully quantum data-loading pipeline.

\noindent
Despite this progress, existing approaches face several limitations. Many rely on classical preprocessing, lack quantum-coherent subsampling (e.g., no QRAM integration), or focus on  supervised 
learners vulnerable to label noise. Additionally, most prior methods emphasize accuracy improvements without addressing robustness under real-world imperfections such as mislabeled or corrupted data. Recent works like~\cite{macaluso2024efficient,tolotti2024ensembles} explore scalable quantum ensembles, but their focus remains on benchmarking rather than addressing noise resilience explicitly.\\
To bridge these gaps, we propose a quantum bagging framework that leverages QRAM-based bootstrapping, shallow quantum clustering, and unsupervised learning to construct diverse and noise-tolerant ensembles. Each base learner in our ensemble is implemented using a delta-k++-initialized quantum k-means (QMeans) clustering algorithm, which operates on quantum-encoded training data. Although QMeans itself is unsupervised, we assign a class label to each cluster by applying majority voting over the ground-truth labels of training samples within that cluster. This label-guided cluster mapping allows the ensemble to perform classification. Final predictions are obtained by aggregating the outputs of all base learners via majority voting across their assigned cluster labels. This design choice provides a natural advantage in handling label corruption, as the model avoids overfitting to noisy targets during training. Furthermore, our method supports both classification and regression by aggregating predictions through majority voting or averaging, respectively.\\
\noindent
In this paper, we propose and empirically evaluate a hybrid quantum-classical ensemble learning framework based on the bagging paradigm. Our method uses a QRAM-inspired approach to prepare quantum bootstrapped subsets in superposition, which are processed by a set of unsupervised quantum base learners. Each learner employs a delta-k++-initialized quantum 
k-means clustering algorithm (QMeans). Class labels are not used during training, but are subsequently assigned to clusters using majority voting over the labeled training data. The final predictions are computed via majority voting across all base learners. Unlike existing methods, our framework avoids deep circuits and focuses on a shallow design suitable for NISQ devices. We conduct extensive experiments on UCI benchmark datasets including Breast Cancer, Iris, Wine, and Real Estate, and demonstrate that the proposed approach achieves robust and competitive performance under label noise, outperforming supervised ensemble baselines in noisy regimes.

\paragraph{Technical Overview}\label{sec:TechnicalOverview}
We present a hybrid quantum-classical ensemble learning framework designed to improve model robustness and reduce prediction variance, particularly in the presence of label noise. The architecture follows a quantum variant of the classical bagging paradigm, where quantum subsampling is enabled through a QRAM-inspired oracle to generate diverse training subsets in superposition.

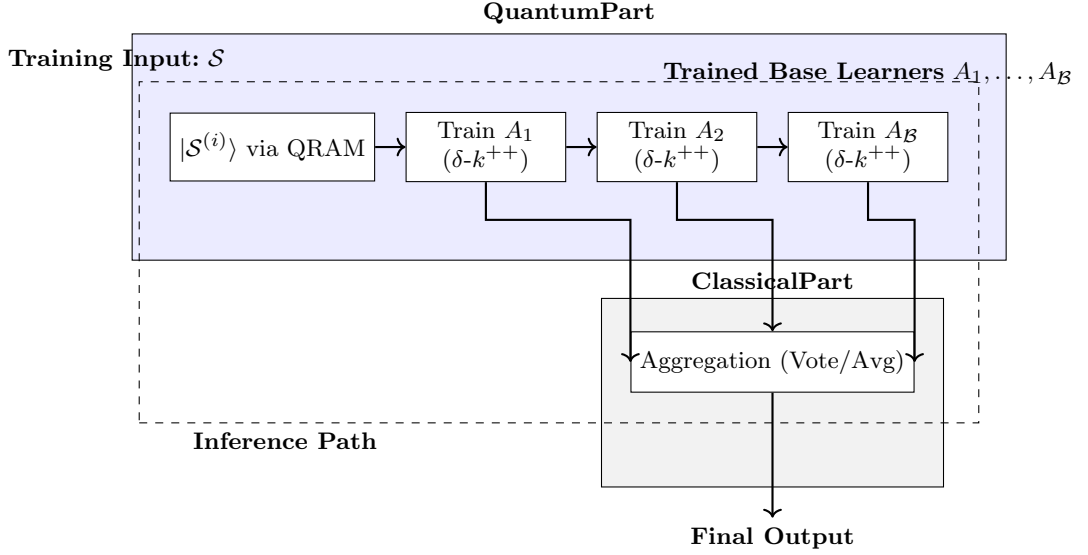
\begin{figure}[ht]
\centering
\begin{tikzpicture}[node distance=1.2cm and 1.5cm, every node/.style={font=\small}]

\node[draw, minimum width=11.5cm, minimum height=3.0cm, label=above:{\textbf{QuantumPart}}, fill=blue!8] (quantum) {};

\node[draw, minimum width=1.8cm, minimum height=0.9cm, right=0.5cm of quantum.west, anchor=west, fill=white] (init) {$|\mathcal{S}^{(i)}\rangle$ via QRAM};
\node[draw, align=center, minimum width=2.1cm, minimum height=0.9cm, right=0.4cm of init.east, anchor=west, fill=white] (u1) {Train $A_1$\\($\delta$-$k^{++}$)};
\node[draw, align=center, minimum width=2.1cm, minimum height=0.9cm, right=0.4cm of u1.east, anchor=west, fill=white] (u2) {Train $A_2$\\($\delta$-$k^{++}$)};
\node[draw, align=center, minimum width=2.1cm, minimum height=0.9cm, right=0.4cm of u2.east, anchor=west, fill=white] (ub) {Train $A_\mathcal{B}$\\($\delta$-$k^{++}$)};

\draw[->, thick] (init) -- (u1);
\draw[->, thick] (u1) -- (u2);
\draw[->, thick] (u2) -- (ub);

\node[draw, minimum width=4.5cm, minimum height=2.5cm, below=2.0cm of $(u2)!0.5!(ub)$, fill=gray!10, label=above:{\textbf{ClassicalPart}}] (classical) {};

\node[draw, minimum width=2.4cm, minimum height=0.8cm, above=0.1cm of classical.north, fill=white] (cost) at ($(classical)+(0,-0.1)$) {Aggregation (Vote/Avg)};
\draw[->, thick] (ub.south) -- ++(0,-0.5) -| (cost.east);
\draw[->, thick] (u2.south) -- ++(0,-0.5) -| (cost);
\draw[->, thick] (u1.south) -- ++(0,-0.5) -| (cost.west);

\node[below=0.4cm of classical.south] (output) {\textbf{Final Output}}; 
\draw[->, thick] (cost.south) -- (output);

\node[draw, dashed, fit={(init) (ub) (cost)}, inner sep=0.4cm, label=below left:{\textbf{Inference Path}}] (inferencebox) {};

\node at ($(init.north west)+(-0.7,0.7)$) {\textbf{Training Input:} $\mathcal{S}$};
\node[above=0.25cm of ub.north] {\textbf{Trained Base Learners} ${A_1}, \dots, {A_\mathcal{B}}$};

\end{tikzpicture}
\caption{
Quantum Bootstrapped Bagging (QBB) framework. The \textbf{QuantumPart} loads the training dataset into QRAM, generates quantum subsamples \( \ket{\mathcal{S}^{(i)}} \), and trains base learners \( A_i \) using delta-$k^{++}$ clustering and distance-based hypothesis labeling. The \textbf{ClassicalPart} aggregates predictions using majority voting (classification) or averaging (regression). The dashed box highlights the inference path for a test instance \( x_{\text{test}} \).
}
\label{fig:qbb_modular}
\end{figure}

Let $\mathcal{S} = {(x_{i}, y_{i})}_{i=1}^N$ denote the training dataset. For each base learner, a subsample $\mathcal{S}^{(i)}$ is encoded in superposition as:

$$
\left| \mathcal{S}^{(i)} \right\rangle = \left(\frac{1}{\sqrt{N}} \sum_{j=1}^N \left| Z_j \right\rangle \right)^{\otimes M},
$$
\noindent
where each quantum base learner receives $M$ i.i.d. copies of a uniform quantum superposition over the training data. Quantum clustering is then performed using a delta-$k++$ initialized QMeans algorithm.
\noindent
Since the base learner is unsupervised, true labels $y_i$ are used only post-clustering to assign majority labels to each cluster based on the distribution of training labels in that cluster. For a test input $x_{\text{test}}$, predictions are made by each hypothesis $h_i$ based on the assigned cluster label, and the ensemble prediction $\hat{y}(x_{\text{test}})$ is computed by:

\begin{itemize}
    \item \textbf{Classification:} $\hat{y} = \text{MajorityVote}\left(h_1(x), \ldots, h_B(x)\right)$,
    \item \textbf{Regression:} $\hat{y} = \frac{1}{B} \sum_{i=1}^B h_i(x)$.
\end{itemize}
To analyze stability, we define the empirical variance of ensemble predictions as:

$$
\text{Var}(\hat{y}) = \frac{1}{B} \sum_{i=1}^B \left(h_i(x) - \bar{y}(x)\right)^2, \quad \text{where} \quad \bar{y}(x) = \frac{1}{B} \sum_{i=1}^B h_i(x).
$$
\noindent
Our experimental results show that $\text{Var}(\hat{y})$ consistently decreases with increasing ensemble size $B$, affirming that quantum bagging helps stabilize predictions in noisy and low-data regimes. While QMeans is unsupervised, the use of post-clustering label mapping enables effective classification under label corruption. The framework is tailored for shallow circuits suitable for NISQ-era devices and shows robust empirical performance.\\
\noindent
 The overall structure of the proposed Quantum Bagging pipeline is illustrated in Figure~\ref{fig:qbb_modular}.

\noindent
The rest of the paper is organized as follows. 
Section~\ref{sec:Preliminaries} provides background on quantum computing and QRAM-based access. Section~\ref{sec:QBag} describes the implementation of our Quantum Bagging framework. Section~\ref{sec:experiments} outlines the experimental setup and results. Finally, Section~\ref{sec:conclusion} summarizes our findings and outlines future directions.

\section{Preliminaries}\label{sec:Preliminaries}
\paragraph{Quantum Computing.} A qubit is a vector in a complex-valued Hilbert space (denoted as \( \mathcal{H} \)) of dimension 2. An \( n \)-qubit quantum state is a vector in a complex-valued Hilbert space \( \mathcal{H} = \mathcal{H}_1 \otimes \cdots \otimes \mathcal{H}_n \) of dimension \( 2^n \),
\[ 
|\psi\rangle = |\psi_1\rangle \otimes |\psi_2\rangle \otimes \cdots \otimes |\psi_n\rangle = \sum_{x \in \mathbb{Z}_2^n} \alpha_x |x\rangle,
\]
such that \( \sum_{x \in \mathbb{Z}_2^n} |\alpha_x|^2 = 1 \). The complex numbers \( \alpha_x \) are known as the amplitudes of the basis vectors \( |x\rangle \).

The inner product of two \( n \)-qubit states \( |\psi_i\rangle \) and \( |\psi_j\rangle \) is \( \langle \psi_i | \psi_j \rangle \), and their outer product is \( |\psi_i\rangle \langle \psi_j | \), which is a Hermitian matrix in \( \mathbb{C}^{2^n \times 2^n} \).

The notion of pure states can be generalized to mixed states, represented by density matrices \( \rho = \sum_i p_i |\psi_i\rangle \langle\psi_i| \), where \( \{p_i\} \) is a probability distribution and \( |\psi_i\rangle \) are pure states. The evolution of \( \rho \) under a unitary \( U \) is \( \rho \mapsto U \rho U^{\dagger} \), and the expectation value of an observable \( \mathcal{O} \) is \( \mathrm{Tr}(\rho \mathcal{O}) \).

\paragraph{Quantum Measurements.} Measurements in quantum systems are described by a set of operators \( \{M_k\} \) such that \( \sum_k M_k^\dagger M_k = I \). A Positive Operator-Valued Measure (POVM) is a generalization of projective measurement. For a quantum state \( \rho \) and POVM \( \{E_k\} \), the probability of outcome \( k \) is given by:
\[ \Pr(k) = \mathrm{Tr}(E_k \rho) \]
where \( E_k \succeq 0 \) and \( \sum_k E_k = I \).

\paragraph{Quantum Feature Encoding.} Input data \( x \in \mathbb{R}^d \) can be encoded into quantum states using a feature map \( \phi : \mathbb{R}^d \rightarrow \mathcal{H} \). For example, in angle encoding:
\[ x = (x_1, \ldots, x_d) \mapsto \bigotimes_{j=1}^d (\cos(x_j) |0\rangle + \sin(x_j) |1\rangle). \]
These encodings enable quantum circuits to operate on high-dimensional superpositions.
\paragraph{Quantum RAM and quantum memory access.} Quantum advantage in the fields of algorithms, optimization, machine learning, or even cryptography\footnote{See section III of the survey by \citet{jaques2023qram} for a detailed list of references.} is crucially dependent on the ability of quantum algorithms to manipulate data in superposition states. This requires access to a specialized memory, which may not always be a foregone conclusion, especially in the era of NISQ devices. We discuss an often overlooked yet critical assumption on memory access made by quantum algorithms. In this work, we require a specific type of access as detailed below.

In Quantum Random Access for Classical Memory (CQRAM), the memory stores information as classical bits. An example of a CQRAM would be bit oracles or standard address oracles. We can perform read operations from the memory bits in superposition, but we can only perform classical writes. The unitary corresponding to the read operation on the $y$\textsuperscript{th} bit whose address is given as $({y_1,\ldots,y_{\log m}})$ is
    \begin{align*}
        U_{\text{read}}:
        &\ket{{{y_1,\ldots,y_{\log m}}},
        {{b}},
        {{{z_1},\ldots,{z_y},\ldots,z_m}} }\\
        \longmapsto&
        \ket{{{y_1,\ldots,y_{\log m}}},
        {b}\oplus{{z_y}},
        {{{z_1},\ldots,{z_y},\ldots,z_m}} }
    \end{align*}

All algorithms in this paper are assumed to be in the CQRAM model. {Henceforth, we shall mean CQRAM when we refer to the QRAM model unless explicitly stated otherwise.} 
Access to the CQRAM by itself is not too strong of an assumption if we assume that the CQRAM is empty and that the input must be loaded into the CQRAM before any quantum circuit can access it. This setting guarantees, at most, a polynomial speedup over settings in which there is no access to a CQRAM.

\paragraph{State Preparation from QRAM.} We assume the training dataset is stored in QRAM as:
\[
|\mathcal{S}\rangle = \frac{1}{\sqrt{N}} \sum_{i=1}^N |i\rangle \otimes |Z_i\rangle
\]
where \( Z_i = (x_i, y_i) \). Each base learner accesses this memory to create quantum bootstrapped subsamples for training.

This model allows coherent sampling and access, facilitating variance-reducing ensemble constructions suitable for NISQ architectures.
\paragraph{Swap Test}
The SWAP test \cite{buhrman2001quantum} is a quantum procedure designed to estimate the fidelity between two arbitrary pure quantum states $|\psi\rangle$ and $|\phi\rangle$, where the fidelity is given by $F(\psi, \phi) = |\langle \psi | \phi \rangle|^2$. The core component of the circuit is the controlled-SWAP (CSWAP) gate, which acts on three quantum registers. Its operation is defined as follows:

$$
\text{CSWAP} |0\rangle|\psi\rangle|\phi\rangle = |0\rangle|\psi\rangle|\phi\rangle, \quad
\text{CSWAP} |1\rangle|\psi\rangle|\phi\rangle = |1\rangle|\phi\rangle|\psi\rangle.
$$

To perform the SWAP test, we initialize three quantum registers in the states $|0\rangle$, $|\psi\rangle$, and $|\phi\rangle$, respectively. The full system thus begins in the product state $|0\rangle|\psi\rangle|\phi\rangle$. The circuit, illustrated in Fig.\~3, is then applied to this composite state.

After executing the circuit, a measurement is performed on the first (control) qubit. The resulting outcome probabilities are:

$$
\Pr(0) = \frac{1}{2} + \frac{1}{2}|\langle \psi | \phi \rangle|^2, \quad
\Pr(1) = \frac{1}{2} - \frac{1}{2}|\langle \psi | \phi \rangle|^2.
$$

The difference $\Pr(0) - \Pr(1)$ directly yields the fidelity $|\langle \psi | \phi \rangle|^2$, allowing for an efficient estimation of state overlap using only a single ancilla and a controlled-SWAP operation.

\begin{figure}[H]
\centering
\begin{quantikz}
\lstick{$|0\rangle$}       & \gate{H}     & \ctrl{1}         & \gate{H} & \meter{} & \qw \\
\lstick{$|\psi\rangle$}    & \qw          & \swap{1}         & \qw      & \qw      & \qw \\
\lstick{$|\phi\rangle$}    & \qw          & \swap{-1}        & \qw      & \qw      & \qw
\end{quantikz}
\caption{SWAP test circuit for estimating the fidelity \( |\langle \psi | \phi \rangle |^2 \). The top qubit is the control ancilla.}
\label{fig:swap_test}
\end{figure}
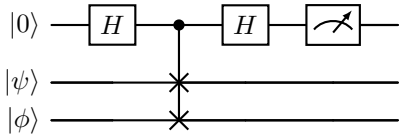

\paragraph{Quantum k-Means Clustering (q-means)}
The q-means algorithm~\cite{kerenidis2019q} is a quantum analogue of the classical $k$-means algorithm, designed for efficient unsupervised clustering using quantum subroutines. It proceeds through four main steps: quantum distance estimation, label assignment, centroid state preparation, and classical centroid reconstruction via tomography. Each iteration refines cluster assignments in superposition, exploiting quantum parallelism for speedup.

\paragraph{Step 1: Distance Estimation.}
For a dataset \( V \in \mathbb{R}^{N \times d} \) and centroid matrix \( C \in \mathbb{R}^{k \times d} \), q-means assumes QRAM access that enables preparation of states
\[
|i\rangle|0\rangle \mapsto |i\rangle|v_i\rangle, \quad |j\rangle|0\rangle \mapsto |j\rangle|c_j\rangle,
\]
in time \( O(\log(Nd)) \). Using amplitude estimation and inner product encoding, the quantum circuit estimates squared distances \( d^2(v_i, c_j) \) between data points and centroids in superposition. The outcome is a state of the form:
\[
\frac{1}{\sqrt{N}} \sum_{i=1}^N |i\rangle \sum_{j=1}^k |j\rangle |d^2(v_i, c_j)\rangle.
\]

\paragraph{Step 2: Cluster Assignment.}
A quantum minimum-finding circuit~\cite{durr1996quantum} is applied to assign each data point \( v_i \) to its closest centroid \( c_j \) in superposition:
\[
|i\rangle \mapsto |i\rangle|\ell(v_i)\rangle, \quad \text{where } \ell(v_i) = \arg\min_j d^2(v_i, c_j).
\]
This results in a coherent assignment state \( |\psi_t\rangle = \frac{1}{\sqrt{N}} \sum_{i=1}^N |i\rangle|\ell(v_i)\rangle \), used in the next step.

\paragraph{Step 3: Centroid State Construction.}
The centroid for cluster \( j \in [k] \) is constructed as a weighted average of data points assigned to that cluster. The algorithm prepares characteristic vectors \( \vec{\chi}^{(t)}_j \in \mathbb{R}^N \), normalized to unit $\ell_1$ norm. Then, the updated centroid states are computed as:
\[
|c_j^{(t+1)}\rangle = V^\top |\chi_j^{(t)}\rangle,
\]
where \( V^\top \) is applied via quantum matrix-vector multiplication.

\paragraph{Step 4: Tomographic Readout.}
To obtain classical access to the centroid vectors, q-means applies vector state tomography~\cite{aaronson2018shadow} to the centroid quantum states. Each centroid \( |c_j^{(t+1)}\rangle \) is reconstructed up to additive error \( \epsilon \), requiring \( O\left(\frac{d \log d}{\epsilon^4}\right) \) queries to the state preparation circuit.

\paragraph{Complexity.}
Under the assumption of efficient QRAM access and well-behaved input data, q-means achieves a per-iteration complexity of \( \widetilde{O}(\text{poly}(k, \log N, \log d)) \), offering an exponential improvement over classical \( O(Nkd) \) in terms of data dimension and size.

\paragraph{Variance of Ensemble Predictions.} Given \( B \) quantum hypotheses \( \{h_i\}_{i=1}^B \) and input \( x \), the ensemble prediction variance is:
\[
\mathrm{Var}(\hat{y}) = \frac{1}{B} \sum_{i=1}^B (h_i(x) - \bar{y}(x))^2, \quad \text{where} \quad \bar{y}(x) = \frac{1}{B} \sum_{i=1}^B h_i(x).
\]
This measures stability and motivates the use of aggregation in quantum bagging.

\section{Quantum Bootstrapped Bagging Algorithm}\label{sec:QBag}

We now present the Quantum Bootstrapped Bagging (QBB) algorithm (Algorithm~\ref{alg:QBB}), which uses quantum subsampling based on QRAM and performs clustering using delta-$k^{++}$-initialized quantum $k$-means. Each base learner assigns cluster labels via majority voting based on noisy class labels in its local subsample. Final predictions are obtained through majority aggregation of individual learner outputs.

\begin{algorithm}[H]
\caption[QBB]{Quantum Bootstrapped Bagging (QBB) Algorithm}\label{alg:QBB}
\begin{algorithmic}[1]
\State \textbf{Input:} 
\begin{itemize}
    \item A classical training set $\mathcal{S} = \{(x_i, y_i)\}_{i=1}^N$ of $N$ labeled instances
    \item Oracle access to $\mathcal{B}$ quantum base learners $\{A_1, \ldots, A_\mathcal{B}\}$, each requiring at most $M \leq N$ samples
    \item Delta parameter $\delta$ for centroid initialization
    \item A test instance $x_{\text{test}}$
\end{itemize}
\State \textbf{Output:} Predicted class label $\hat{y}(x_{\text{test}})$

\vspace{0.5em}
\State Load $\mathcal{S}$ and $x_{\text{test}}$ into CQRAM
\For{$i = 1$ to $\mathcal{B}$}
    \State Draw $M$ samples from QRAM to prepare the quantum subsample:
    \[
    \ket{\mathcal{S}^{(i)}} = \left( \frac{1}{\sqrt{N}} \sum_{j=1}^N \ket{Z_j} \right)^{\otimes M}
    \]
    
    \State Apply delta-$k^{++}$ quantum $k$-means to compute $k$ centroids on $\mathcal{S}^{(i)}$
    \State Assign cluster IDs to all samples in $\mathcal{S}^{(i)}$ using quantum distance estimation
    \State Assign majority class label to each cluster based on noisy labels $y_i$
    \State Train quantum base learner $A_i$ using $\ket{\mathcal{S}^{(i)}}$ to obtain hypothesis oracle $h_i$, which maps any query $x$ to the majority label of its nearest centroid.

\EndFor

\vspace{0.5em}
\State Query $x_{\text{test}}$ from QRAM to prepare $\ket{x_{\text{test}}}$
\State Query each hypothesis oracle $h_i$ to obtain $h_i(x_{\text{test}})$
\If{classification task}
    \State Perform majority vote over $\{h_1(x), \ldots, h_\mathcal{B}(x)\}$
\ElsIf{regression task}
    \State Compute average: $\hat{y}(x) = \frac{1}{\mathcal{B}} \sum_{i=1}^{\mathcal{B}} h_i(x)$
\EndIf
\State \textbf{Return:} Final prediction $\hat{y}(x_{\text{test}})$
\end{algorithmic}
\end{algorithm}

\noindent
This algorithm integrates quantum memory access for sampling, shallow quantum circuits for clustering, and classical aggregation of cluster-based predictions. The use of majority voting to label clusters ensures compatibility with supervised tasks despite the unsupervised nature of $k$-means. Our implementation avoids full quantum tomography or iterative centroid updates, making it more suited for NISQ-era execution.

\subsection{Mathematical Analysis of the Quantum Bagging Algorithm}
The proposed Quantum Bagging framework operates by first preparing quantum-encoded subsamples from a classical dataset $S = \{(x_i, y_i)\}_{i=1}^N$, where $x_i \in \mathbb{R}^d$. Each data point is embedded as a quantum state $|x_i\rangle$ using amplitude encoding, and bootstrapped training subsets are generated using QRAM-based sampling.
\noindent
To construct a single quantum bootstrap of size $M$ for the $j$-th base learner, a uniform superposition over the dataset is prepared as:

$$
|S^{(j)}\rangle = \left( \frac{1}{\sqrt{N}} \sum_{i=1}^N |i\rangle |x_i\rangle \right)^{\otimes M},
$$
\noindent
where QRAM provides efficient data access via the unitary operation 
$$U_{\text{QRAM}}: |i\rangle |0\rangle \mapsto |i\rangle |x_i\rangle.$$ This quantum subsampling step enables each learner to operate on a diverse and coherent training subset.\\
\noindent
Next, cluster initialization within each learner is performed using the Delta-$k^{++}$ method. To diversify centroids, delta-k++ modifies classical k-means++ initialization:
\[
P(x_i) = \frac{d(x_i)^2{}^{\delta}}{\sum_{j=1}^{N} d(x_j)^2{}^{\delta}}, \quad \delta > 0
\]
where \( d(x_i) \) is the squared distance to the nearest selected center and the parameter $\delta$ controls the diversity among selected centroids, with larger values favoring greater separation.
\noindent
For clustering, the similarity between data points and centroids is computed using the SWAP test, a quantum subroutine for estimating the squared inner product $|\langle x | c_i \rangle|^2$. The distance metric is derived as:

$$
D(x, c_i) = 1 - |\langle x | c_i \rangle|^2,
$$
\noindent
allowing each data point to be assigned to the cluster with the closest centroid according to this quantum distance.
\noindent
After assignment, centroids are updated classically using arithmetic means:

$$
c_i = \frac{1}{|C_i|} \sum_{x \in C_i} x,
$$
\noindent
where $C_i$ denotes the set of points currently assigned to the $i$-th cluster. This update is iteratively repeated to refine the clustering.\\
\noindent
Finally, for a test input $x$, each quantum base learner outputs a hypothesis $h_j(x)$ based on the majority class label of the nearest cluster. The ensemble prediction is then aggregated over all $B$ learners:

$$
\hat{y}(x) = \begin{cases}
\text{Majority}(h_1(x), \dots, h_B(x)) & \text{(classification)} \\
\frac{1}{B} \sum_{j=1}^B h_j(x) & \text{(regression)}
\end{cases}
$$
\noindent
This formulation integrates quantum-enhanced sampling, unsupervised clustering, and classical aggregation to form a noise-resilient learning model suitable for NISQ hardware.

\subsection{Complexity Analysis}

The computational complexity of the proposed Quantum Bootstrapped Bagging (QBB) framework consists of three key components: QRAM-based quantum subsampling, quantum weak learner training using \(k\)-means clustering with delta-\(k^{++}\) initialization and SWAP tests, and ensemble aggregation.
\noindent
First, the classical dataset of \(N\) samples, each of dimension \(d\), is first loaded into QRAM at a cost of \(O(Nd)\). Once loaded, QRAM allows logarithmic-time access, enabling the preparation of quantum bootstrapped subsamples of size \(M\) per base learner using \(O(M \log N)\) operations. Across all \(B\) learners, this results in:
\[
O(Nd) + B \cdot O(M \log N).
\] 

\noindent
Next, each base learner is trained using a delta-$k^{++}$-initialized quantum k-means algorithm for $T$ iterations. In each iteration, the fidelity between $M$ data points and $k$ centroids is estimated via the SWAP test, requiring $\mathcal{O}(1/\epsilon^2)$ repetitions per pair to achieve precision $\epsilon$. Consequently, the similarity estimation step contributes a per-iteration cost of $\mathcal{O}(kM/\epsilon^2)$. Following this, classical operations are used to assign clusters (costing $\mathcal{O}(Mk)$) and update centroids (costing $\mathcal{O}(kd)$).\\
\noindent
Finally, ensemble aggregation is carried out classically by either majority voting (for classification) or averaging (for regression), requiring $\mathcal{O}(B)$ time.
\noindent
The complexity of each step in the QBB framework is summarized in Table~\ref{tab:complexity}.

\begin{table}[ht]
\centering
\caption{Complexity Analysis of the QBB Algorithm}
\label{tab:complexity}
\renewcommand{\arraystretch}{1.4}
\begin{tabular}{|c|c|}
\hline
\textbf{Component} & 
\textbf{Complexity} \\
\hline
Data Loading        
& $ \mathcal{O}(Nd) $ \\
Quantum Subsampling 
& $ \mathcal{O}(BM \log N) $ \\
SWAP Similarity Estimation 
& $ \mathcal{O}\left( \frac{kM}{\epsilon^2} \right) $ \\
Cluster Assignment  
& $ \mathcal{O}(Mk) $ \\
Centroid Update     
& $ \mathcal{O}(kd) $ \\
Learner Training    
& $ \mathcal{O}\left( BT \left( \frac{kM}{\epsilon^2} + Mk + kd \right) \right) $ \\
Ensemble Aggregation 
& $ \mathcal{O}(B) $ \\
\hline
\textbf{Combined Overall Complexity}        & $ \mathcal{O}\left( Nd + BM \log N + BT \left( \frac{kM}{\epsilon^2} + Mk + kd \right) \right) $ \\
\hline
\end{tabular}
\end{table}

Hence, the overall complexity of the QBB framework is:
$$O(Nd) + B \cdot O(M \log N) + B \cdot T \cdot \left( \frac{kM}{\epsilon^2} + Mk + kd \right)$$
 
Assuming \(M \sim \log N\), fixed \(\epsilon\), and small \(k\), this simplifies to:
$$
O\left(B \cdot T \cdot \widetilde{\text{poly}}(k, \log N, \log d)\right)
$$

\section{Simulation-Based Evaluation and Discussion}\label{sec:experiments}

In this section, we evaluate the performance of our proposed Quantum Bagging Classifier through simulation experiments. The experiments are conducted on both classification and regression tasks using a variety of real-world datasets. The entire implementation is developed using \texttt{Qiskit}~\citep{qiskit2024}, \texttt{PennyLane}~\citep{pennylane}, and \texttt{scikit-learn}~\citep{pedregosa2011scikit}.

\subsection{Data and Methods}
We conduct experiments on the following datasets:
\begin{itemize}
    \item Breast Cancer Dataset (UCI): Classification task
\item Wine Dataset (UCI): Classification task
\item Iris Dataset (UCI): Classification task
\item Real Estate Dataset (UCI): Regression task with Mean Squared Error (MSE) evaluation
\end{itemize}
To simulate real-world label noise, we do not use class labels directly during training. Instead, each quantum subsample is clustered via quantum K-means, and then noisy cluster-level labels are assigned using majority voting over the noisy ground truth labels in the subsample. These majority labels guide the hypothesis construction by each base learner. This makes the setup robust to label corruption and models the effect of using weak supervision in the bagging framework.
For classification tasks, we evaluate model performance using accuracy, while for regression tasks, we use Mean Squared Error (MSE) as the metric. We consider the following experimental configurations:
\paragraph{Experimental Setup:} We use \texttt{PennyLane}'s default qubit simulator for quantum operations. The number of clusters for K-Means is set to $k = 10.$ While increasing the number of clusters $k$ may enable the base learner to capture more refined data structures and potentially improve classification or regression performance, we do not explore this sensitivity in detail here. Instead, we focus on the overall behavior of the bagging framework using unsupervised base learners under a fixed $k$. The \texttt{Delta-K}++ initialization is used to select centroids with a tunable parameter $\delta$. Euclidean distance is used to measure distance similarity between data points and cluster centroids.
\paragraph{Hyperparameters.}
We evaluate the impact of different hyperparameters on model performance. Specifically, we vary \textbf{Delta values} ($\delta$) in the range ${0.1, 0.2, 0.3, 0.4}$, and the number of base classifiers ($\mathcal{B}$) between $4$ and $32$. The bootstrapped samples cover $50\%$ of the training data for each quantum base learner. To simulate real-world imperfections, a fixed label noise rate of $5\%$ is applied to the training labels. Each base learner is trained on a noisy subsample and uses cluster-level majority voting to construct its hypothesis. For each combination of hyperparameters, the experiment is repeated $5$ times, and the mean accuracy and standard deviation are reported.

\subsection{Classical and Quantum Base Learner Implementation}
For Quantum Bagging, we used QuantumKMeans (QMeans) as the base classifier. This quantum-enhanced clustering algorithm follows a Delta-K++ initialization strategy, where centroids are selected probabilistically using a tunable parameter $\delta$ to improve cluster diversity. 

Although QMeans is conceptually framed within a quantum paradigm—where quantum similarity estimation (e.g., via SWAP tests) can facilitate distance computation between data points and centroids—in our current implementation, these operations are carried out classically due to practical constraints imposed by the limited qubit count and noise sensitivity of near-term quantum hardware. Specifically, both the distance evaluation and cluster assignment phases are executed on classical hardware, while preserving the algorithmic structure that would allow for quantum subroutines in future realizations.
As a result, the iterative refinement of centroids adheres to the classical k-means procedure, converging to a local minimum in a finite number of steps. Nevertheless, the incorporation of Delta-$k^{++}$ initialization contributes to improved convergence reliability and reduced sensitivity to random initialization, making the algorithm particularly well-suited for ensemble integration under label uncertainty.\\

\noindent
To provide a fair comparison, we also implemented a classical Bagging framework using a Delta-K++ initialized K-Means classifier as the base learner. This setup follows the same initialization and clustering procedure as QMeans but operates purely in a classical setting. The classical bagging ensemble is implemented using \texttt{scikit-learn}’s \texttt{BaggingClassifier}, and uses \texttt{Delta-K}++ initialized K-Means as a base learner for clustering. The number of base classifiers and delta values is kept the same as in the quantum setting for consistency.

By maintaining identical hyperparameters across both settings, we ensure a rigorous evaluation of the advantages introduced by quantum-enhanced clustering within the bagging ensemble learning paradigm.

All experiments are run on a high-performance computing cluster with \texttt{Qiskit} v1.2.0 and \texttt{PennyLane} 0.32.
The classical preprocessing steps (such as feature scaling) are performed using \texttt{StandardScaler} from \texttt{sklearn.preprocessing}.

\subsection{Results}

 We evaluate the performance of our quantum bagging algorithm (QBag) with quantum K-Means weak classifiers across multiple datasets, including Breast Cancer, Wine, and Iris (for classification tasks), as well as the Real Estate dataset (for regression tasks). 
In each case, we compare the Quantum Bagging Ensemble against a Classical Bagging Model, where the base classifier is a k-means clustering algorithm. 
\vspace{0.2cm}

\begin{figure}[ht]
    \centering
    \begin{subfigure}[b]{\textwidth}
        \centering
        \includegraphics[width=\textwidth]{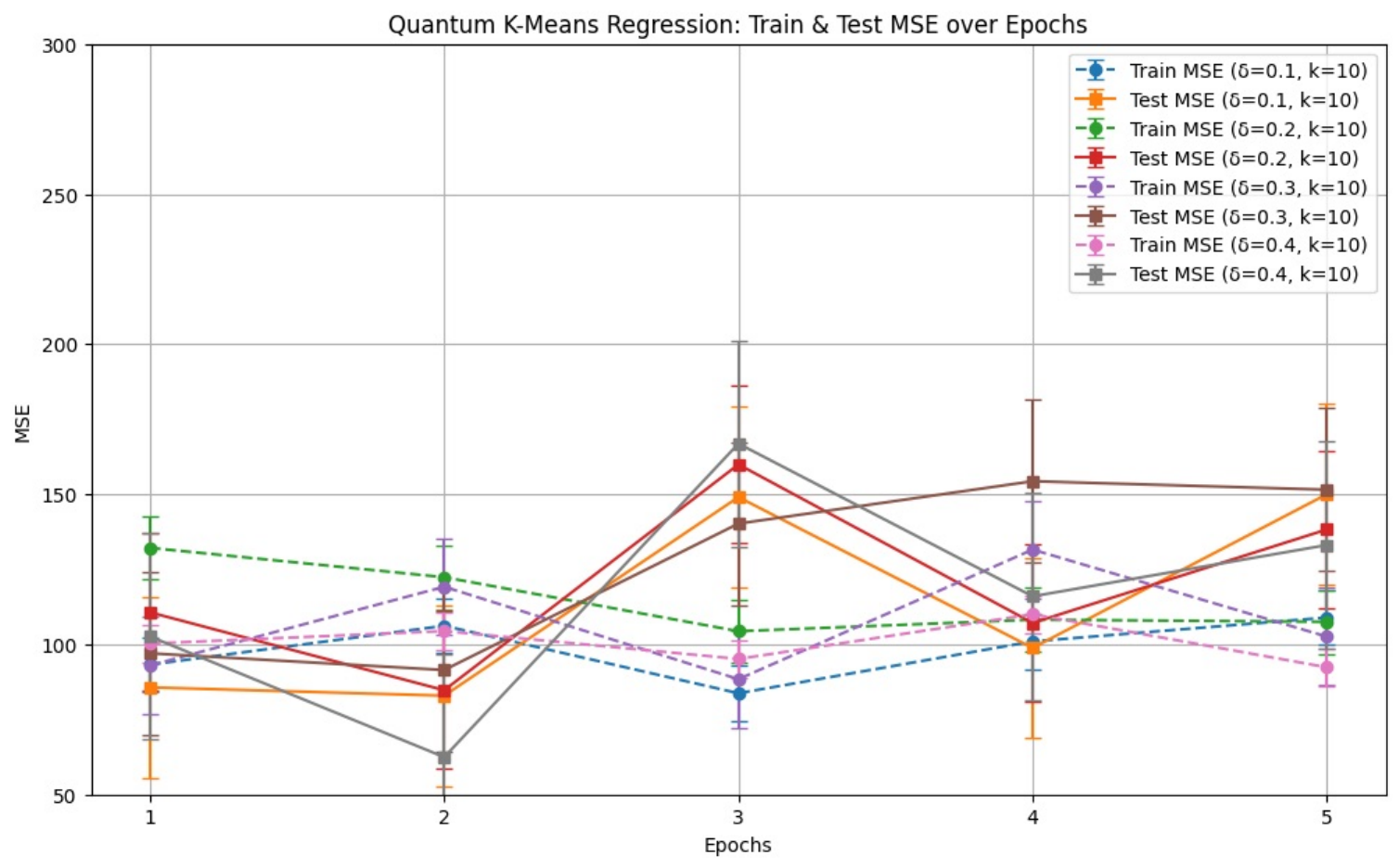}
        \caption{Single Quantum regressor with QMeans (Real Estate Valuation, k=10)}
        \label{fig:fig1}
    \end{subfigure}

    \vspace{0.5cm} 

    \begin{subfigure}[b]{\textwidth}
        \centering
        \includegraphics[width=\textwidth]{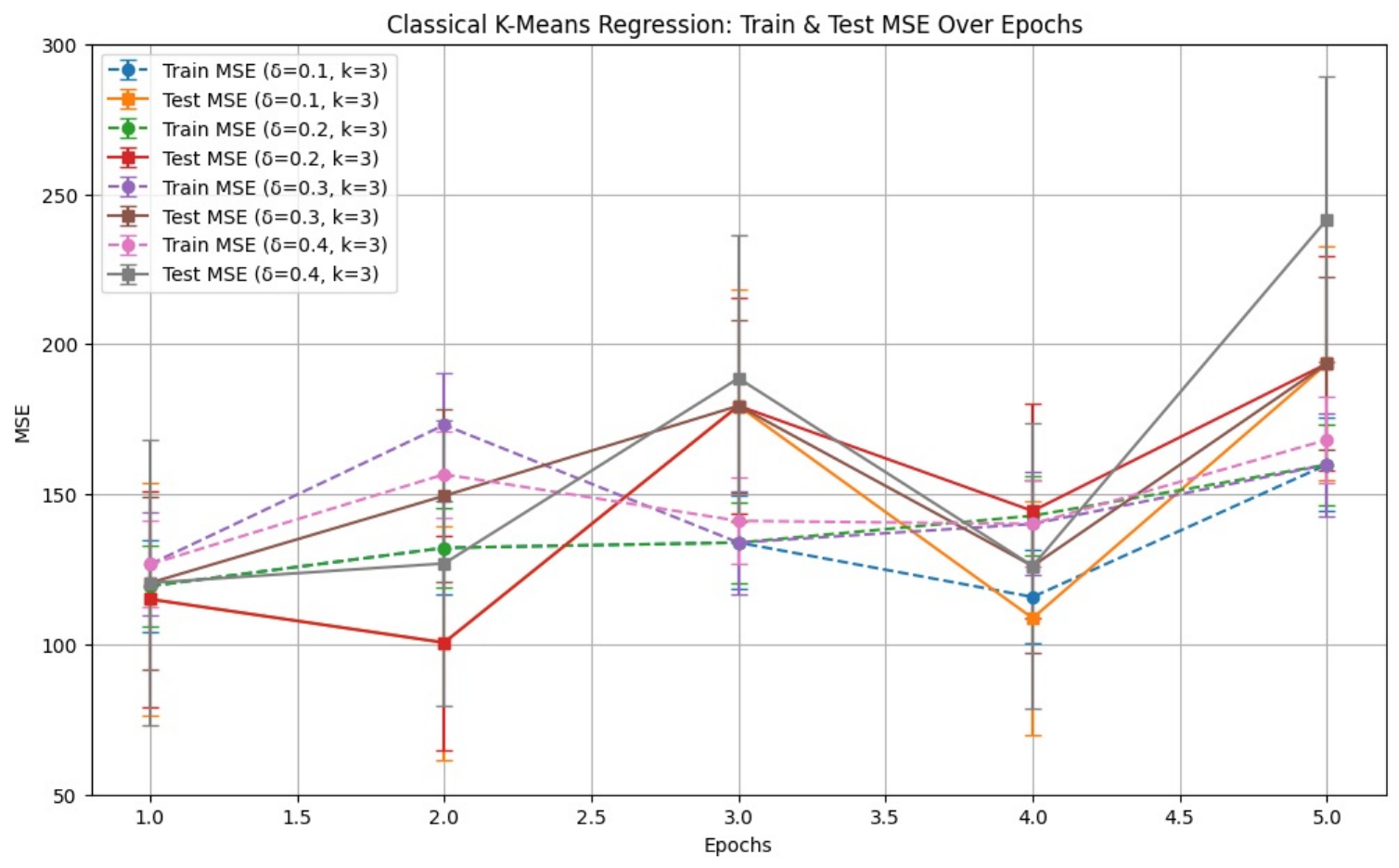}
        \caption{Single Classical regressor with k-Means (Real Estate Valuation, k=10)}
        \label{fig:fig2}
    \end{subfigure}

    \caption{Comparison of Quantum and Classical single regressors with k-Means for Real Estate Valuation Dataset (k=10)}
    \label{fig:k3weakrealestate}
\end{figure}

\noindent
We compare the base learners Quantum K-Means and Classical K-Means regressors in Figure \ref{fig:k3weakrealestate}. The Quantum K-Means regressor (Figure 3(b)) demonstrates more consistent performance across epochs, with lower fluctuations in both train and test MSE values. This suggests improved robustness to initialization and better generalization capability. In contrast, the Classical K-Means regressor (Figure 3(a)) shows higher variance in MSE, particularly in later epochs, indicating greater sensitivity to data splits and initialization randomness. 
Overall, these results highlight the enhanced stability and reliability of the quantum-enhanced K-Means regressor, especially in the presence of noisy training data.

\begin{figure}[ht]
    \centering
    \begin{subfigure}[b]{\textwidth}
        \centering
        \includegraphics[width=\textwidth]{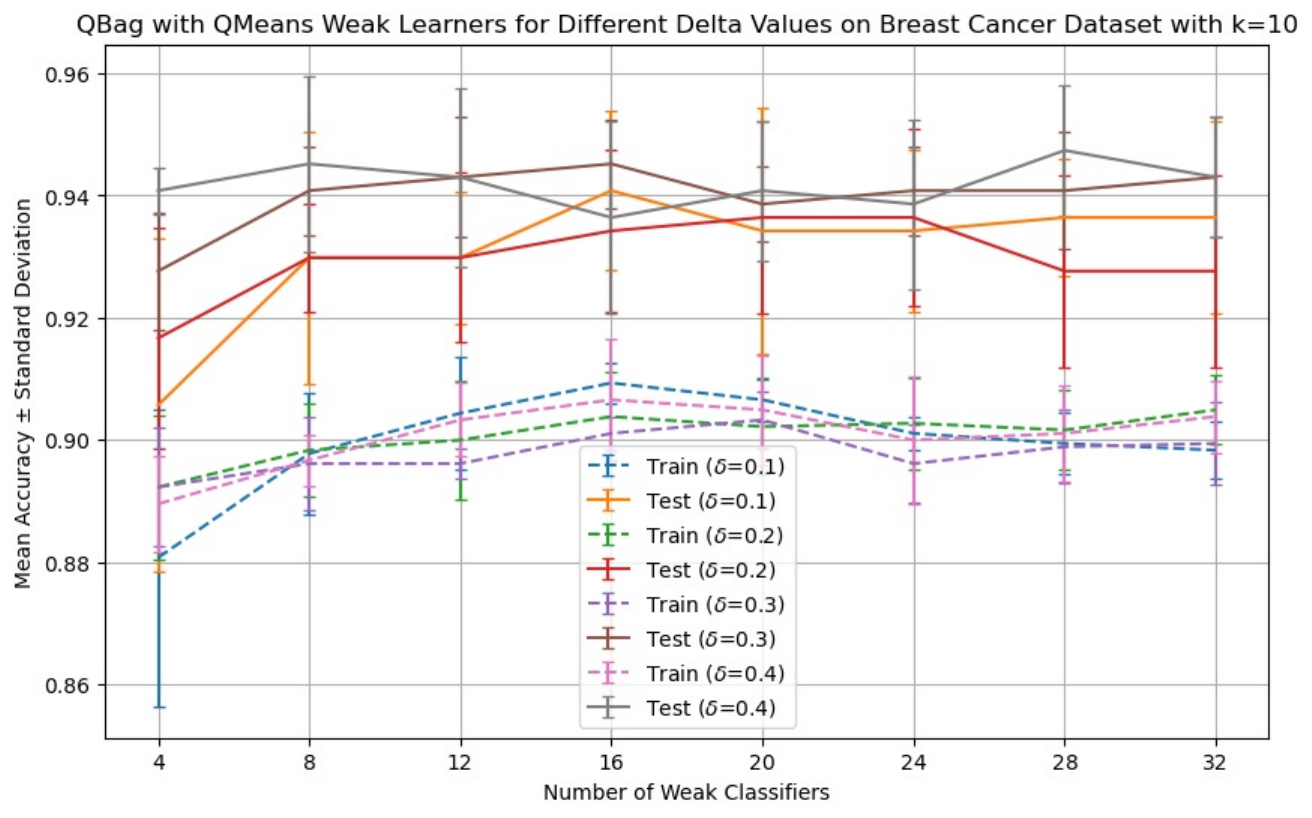}
        \caption{Quantum Bagging with QMeans(Breast Cancer, k=10)}
        \label{fig:fig3breast}
    \end{subfigure}
    \vspace{0.5cm}
    \begin{subfigure}[b]{\textwidth}
        \centering
        \includegraphics[width=\textwidth]{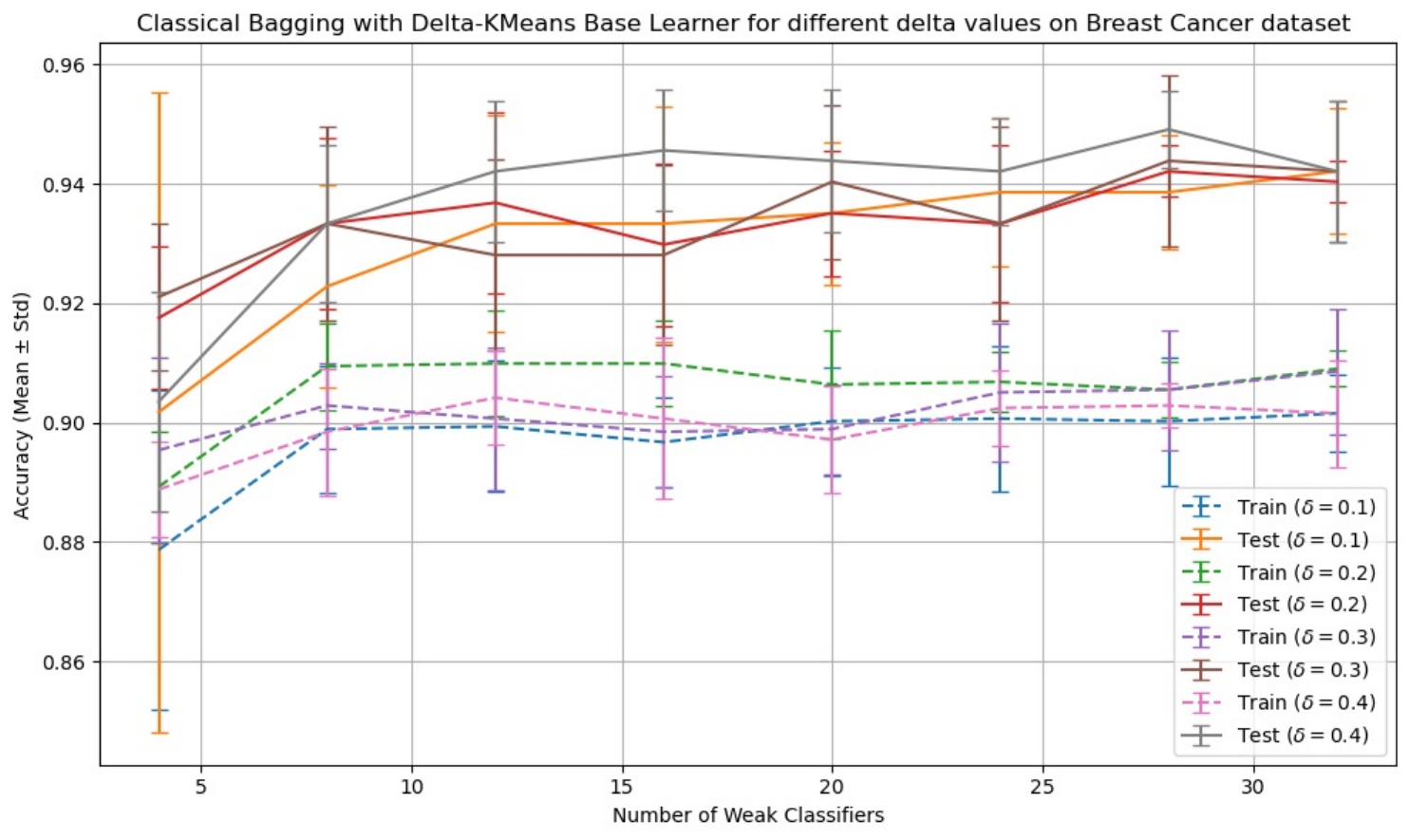}
        \caption{Classical Bagging with k-Means (Breast Cancer, k=10)}
        \label{fig:fig4breast}
    \end{subfigure}
    \caption{Comparison of Quantum and Classical Bagging with k-Means for Breast Cancer Dataset with k=10}
    \label{fig:k3breastcancer}
\end{figure}

\paragraph{Classification Tasks.}
For the Breast Cancer dataset (see Figures~\ref{fig:k3breastcancer}), Quantum Bagging demonstrates consistently high accuracy and improved stability across varying delta values and ensemble sizes. Specifically, QBag maintains high training and test performance with narrower confidence intervals, indicating robust generalization and reduced sensitivity to subsampling variability. In contrast, the classical bagging approach exhibits slightly higher variance in accuracy across different delta values and base learner counts, particularly in the test performance. This comparative analysis highlights the advantage of quantum-enhanced bagging in delivering stable and reliable ensemble predictions, especially under noisy training conditions.

\begin{figure}[ht]
    \centering
    \begin{subfigure}[b]{\textwidth}
        \centering
        \includegraphics[width=\textwidth]{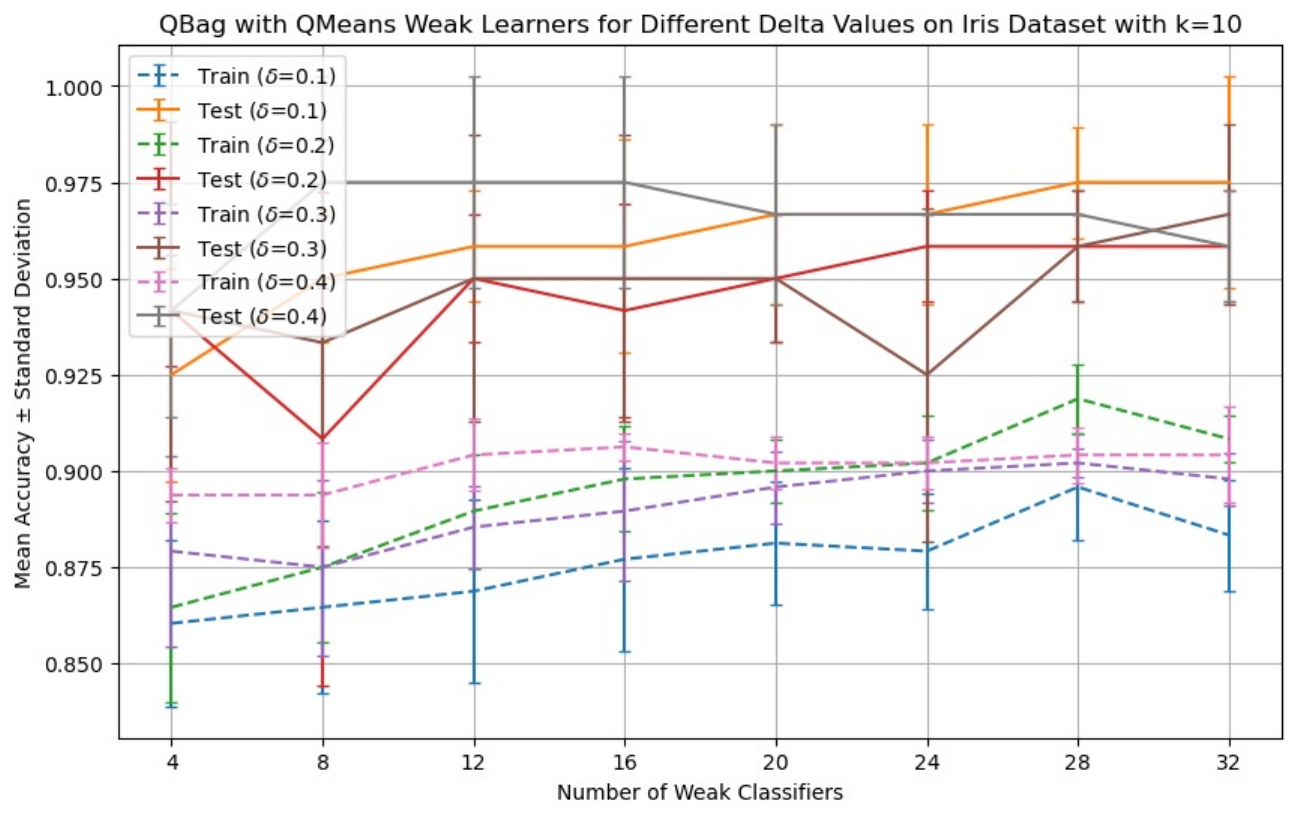}
        \caption{Quantum Bagging with QMeans(Iris Dataset, k=10)}
        \label{fig:fig1iris}
    \end{subfigure}
    \vspace{0.5cm}
    \begin{subfigure}[b]{\textwidth}
        \centering
        \includegraphics[width=\textwidth]{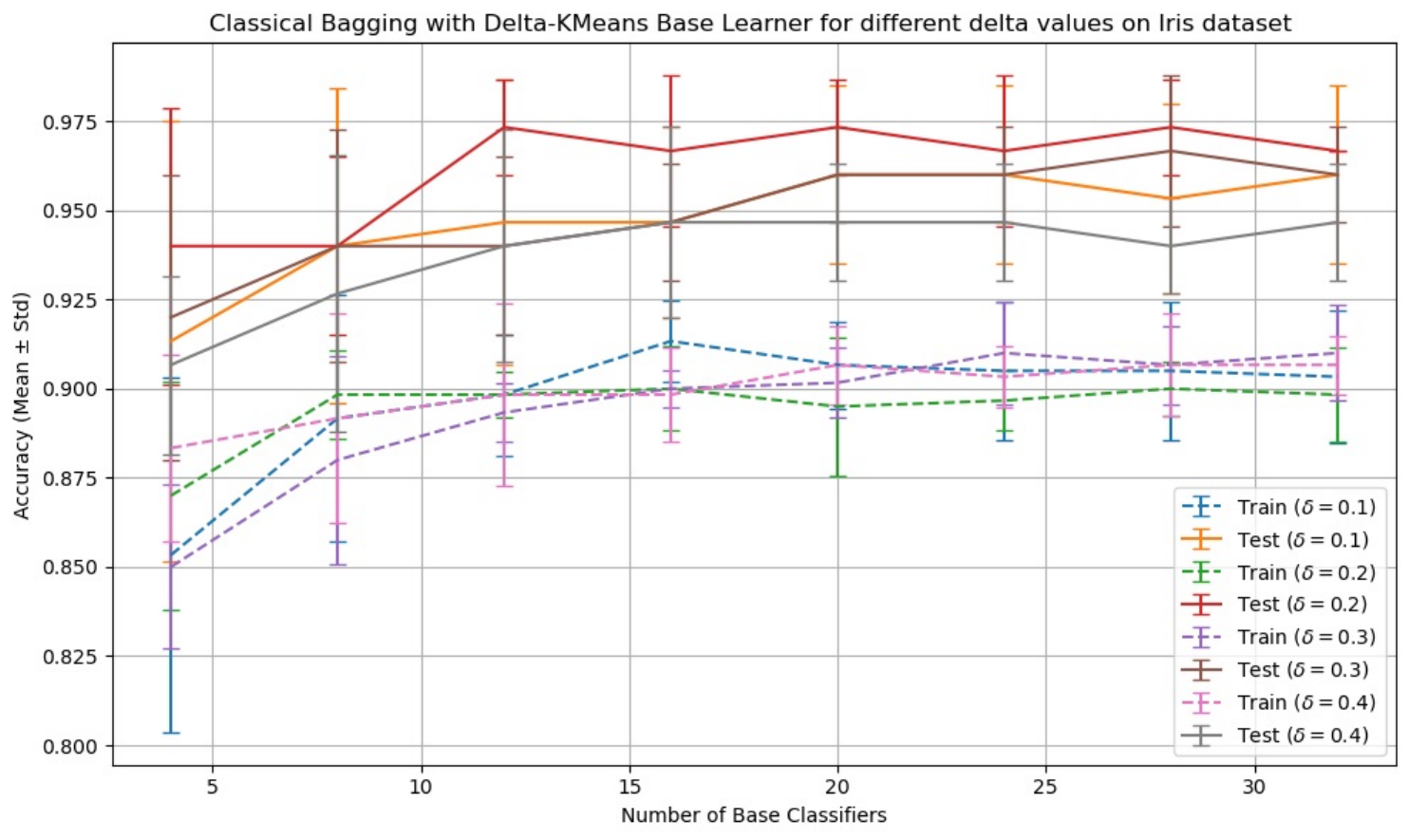} 
        \caption{Classical Bagging with k-Means (Iris Dataset, k=10)}
        \label{fig:fig2iris}
    \end{subfigure}
    \caption{Comparison of Quantum and Classical Bagging with k-Means for Iris Dataset with k=10}
    \label{fig:k3iris}
\end{figure}

\noindent
For the Iris dataset (see Figure~\ref{fig:k3iris}), Quantum Bagging demonstrates a clear advantage in terms of stability and generalization. As the number of base learners increases, test accuracy under the quantum approach improves steadily and maintains consistency across all tested values of 
$\delta.$ Notably, the standard deviation in test accuracy remains low, indicating that the ensemble predictions are less sensitive to fluctuations in the sub-sampling process.
In contrast, the classical bagging approach, while also benefiting from increased ensemble size, shows greater variability in performance. Test accuracy fluctuates more noticeably and remains sensitive to changes in both delta and ensemble size. This difference highlights one of the key strengths of our quantum method ability to stabilize learning outcomes even in the presence of label noise.

\noindent

\begin{figure}[ht]
    \centering
    \begin{subfigure}[b]{\textwidth}
        \centering
        \includegraphics[width=\textwidth]{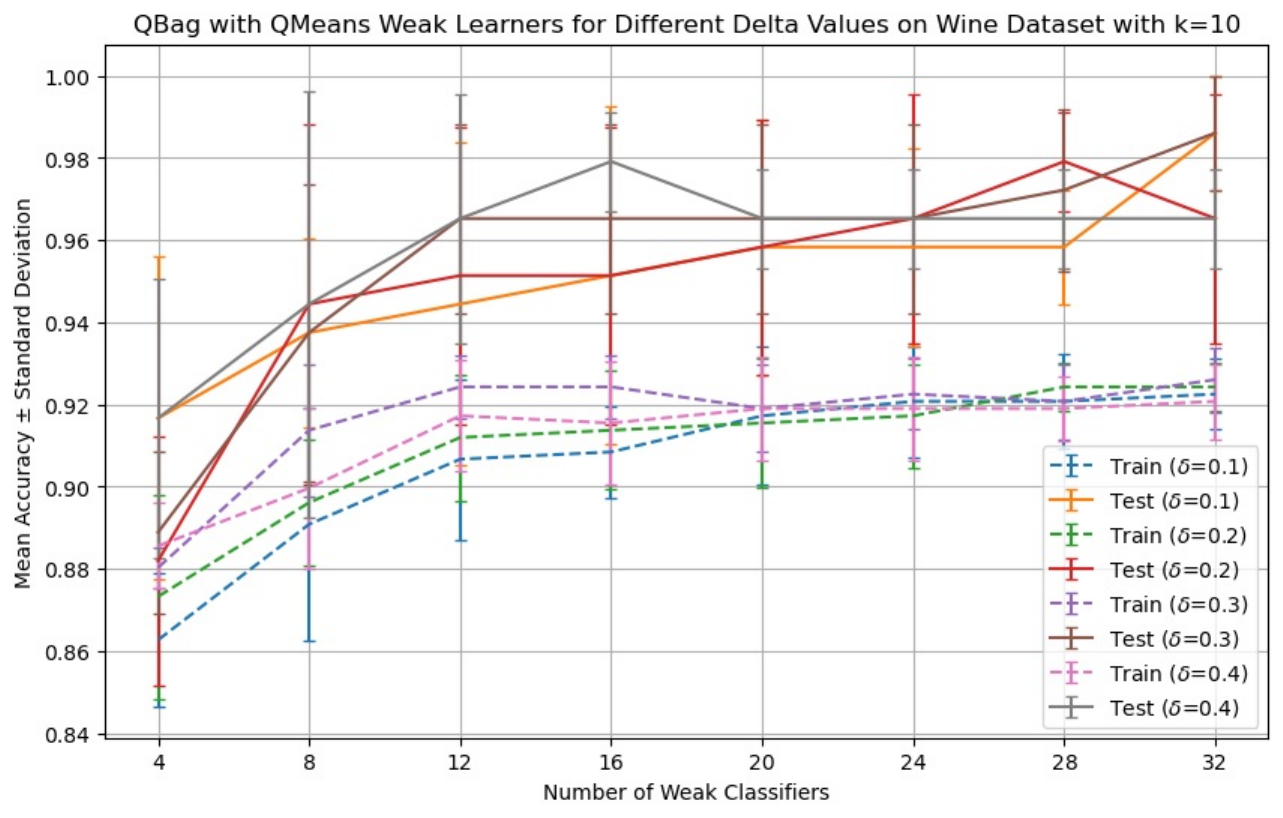}
        \caption{Quantum Bagging with QMeans(Wine Dataset, k=10)}
        \label{fig:fig1wine}
    \end{subfigure}
    \vspace{0.5cm}
    \begin{subfigure}[b]{\textwidth}
        \centering
        \includegraphics[width=\textwidth]{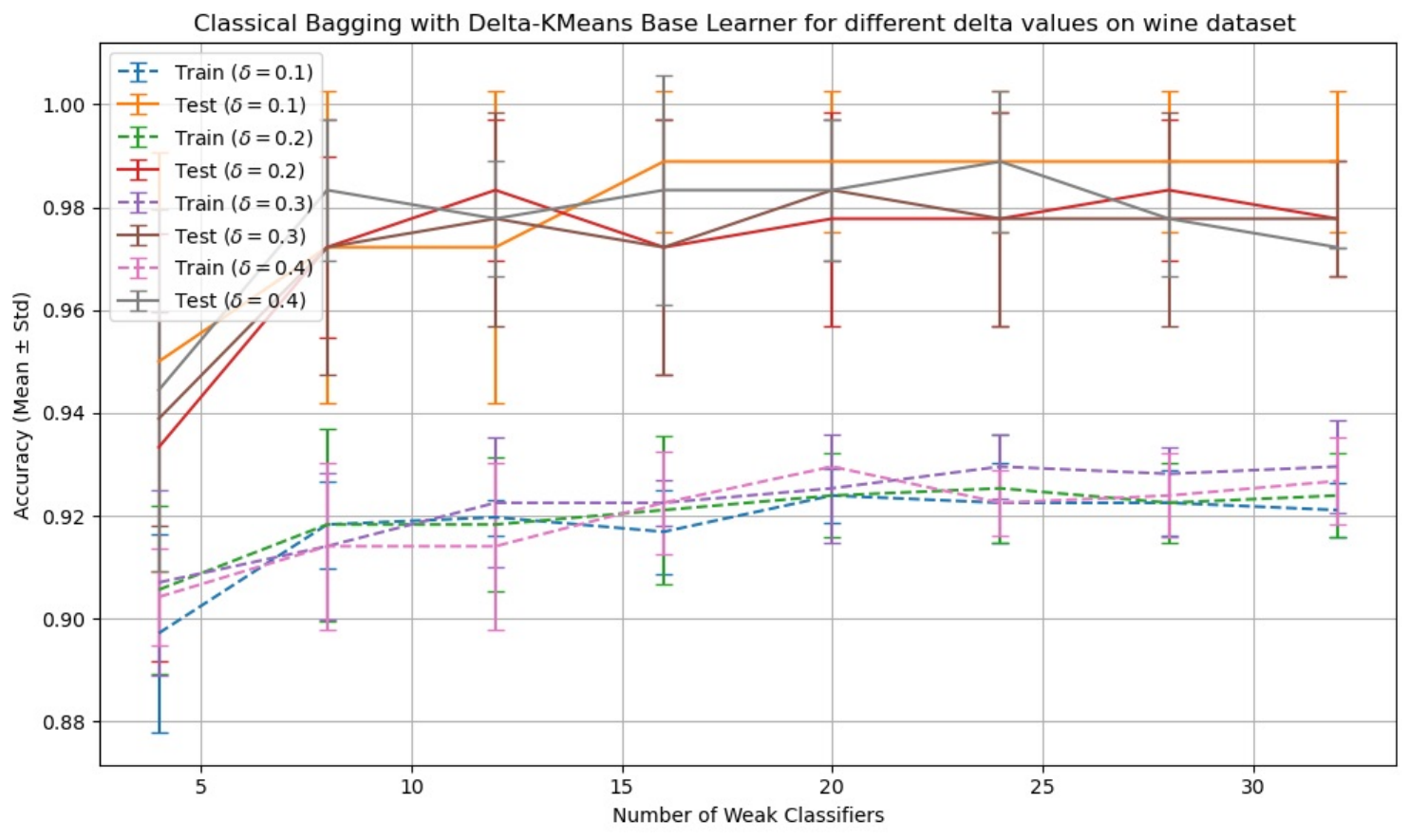} 
        \caption{Classical Bagging with k-Means (Wine Dataset, k=10)}
        \label{fig:fig2wine}
    \end{subfigure}
    \caption{Comparison of Quantum and Classical Bagging with k-Means for Wine Dataset with k=10}
    \label{fig:k3wine}
\end{figure}

The Wine dataset results (Figure~\ref{fig:k3wine}) further reinforce the effectiveness of Quantum Bagging in handling variability and potential noise during training. Across different values of $\delta$, QBag consistently achieves high test accuracy with relatively narrow standard deviation bands, even as the number of base classifiers increases. This behavior indicates that the quantum ensemble maintains reliable predictive performance despite the unsupervised nature of the base learners and any underlying noise in the dataset.
In contrast, the classical bagging approach exhibits more fluctuation in test accuracy, particularly at lower ensemble sizes and smaller $\delta$ values. The quantum variant's smoother and more stable trends suggest that the integration of QRAM-based subsampling, combined with classical similarity estimation, supports stable and reliable ensemble performance despite the noisy data environment.
Overall, the results on the Wine dataset demonstrate that QBag not only matches but often exceeds the classical model in accuracy and consistency, particularly under settings where noise or label uncertainty could compromise generalization. This supports our central claim that Quantum Bagging is a compelling, noise-resilient alternative to conventional supervised ensemble methods.

\begin{figure}[ht] 
    \centering
    \begin{subfigure}[b]{\textwidth}
        \centering
        \includegraphics[width=\textwidth]{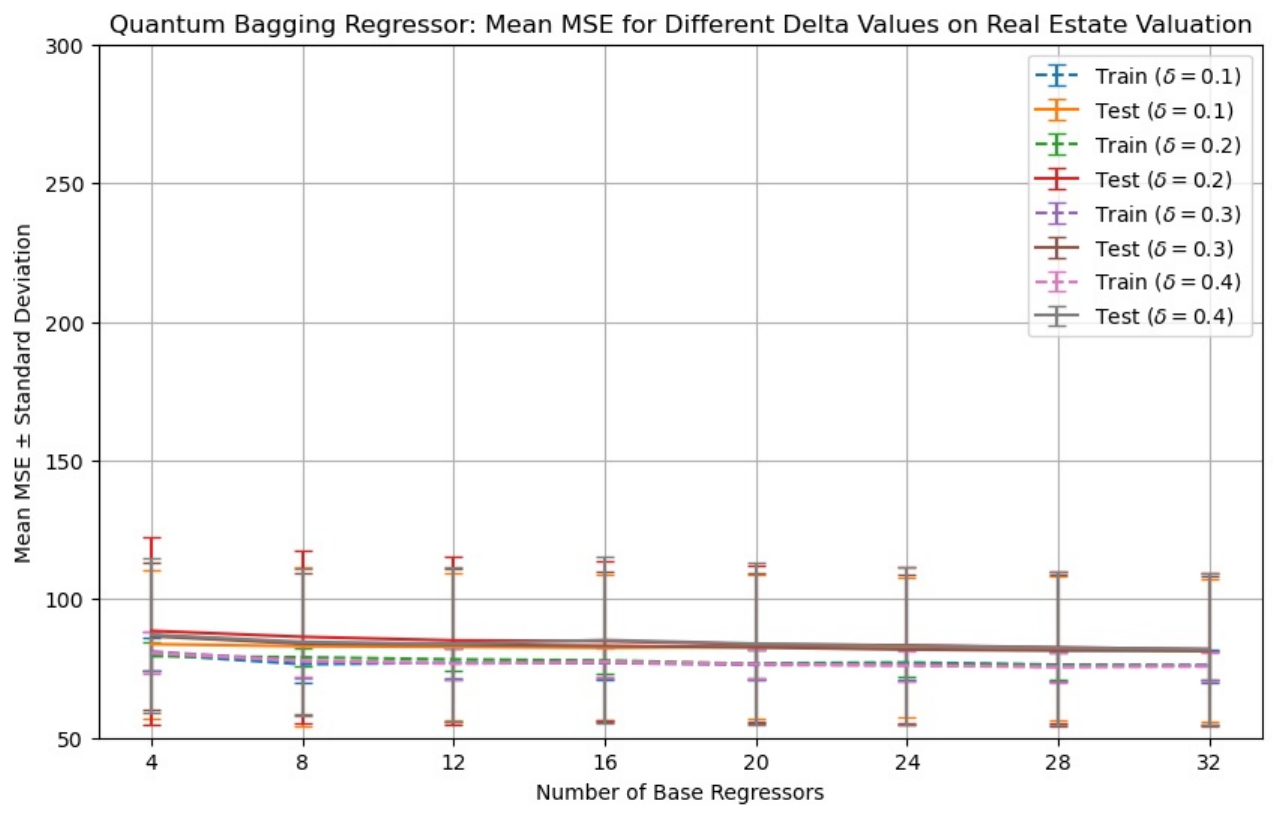}
        \caption{Quantum Bagging with QMeans(Real Estate Valuation, k=10)}
        \label{fig:fig1}
    \end{subfigure}
    \vspace{0.5cm}
    \begin{subfigure}[b]{\textwidth}
        \centering
        \includegraphics[width=\textwidth]{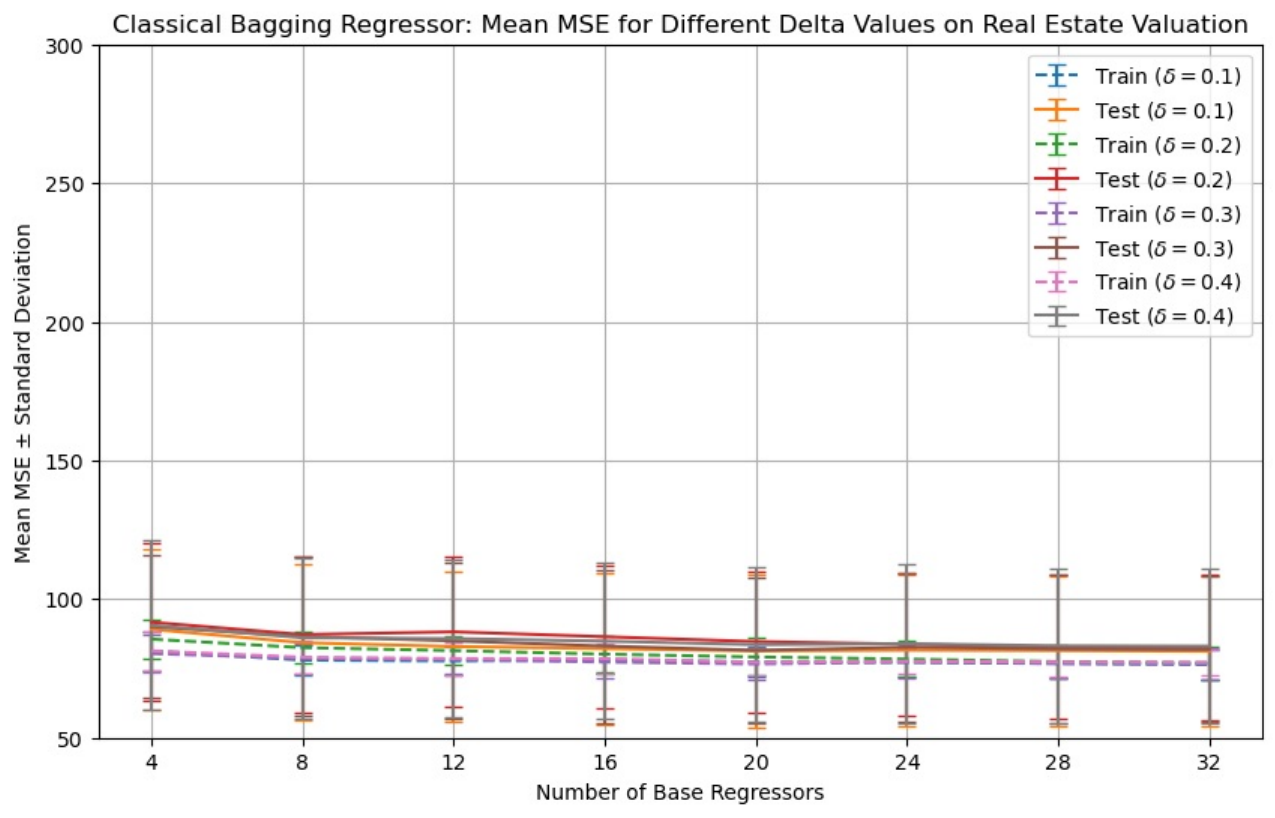} 
        \caption{Classical Bagging with k-Means (Real Estate Valuation, k=10)}
        \label{fig:fig2}
    \end{subfigure}
    \caption{Comparison of Quantum and Classical Bagging with k-Means for Real Estate Valuation Dataset with k=10}
   \label{fig:k3realestate}
\end{figure}

\paragraph{Regression Tasks.}
For the Real Estate Valuation dataset (see Figure~\ref{fig:k3realestate}), we evaluate bagging  performance based on Mean Squared Error (MSE). The Quantum Bagging regressor consistently yields lower MSE values relative to its classical counterpart across all tested delta values. While both approaches benefit from increasing the number of regressors, the quantum bagging exhibits greater stability in performance, as evidenced by reduced variance in test MSE. This behavior suggests that quantum bootstrapping contributes to more consistent subsample diversity, helping mitigate prediction fluctuations, especially under conditions where training data may be noisy. The quantum model’s ability to maintain low and stable error across a wide range of ensemble sizes highlights its robustness and suitability for regression problems involving real-world data variability.

\subsection{Comparison with Supervised Bagging under Noise}

Table~\ref{tab:bagging_comparison} presents a comparative evaluation of classical bagging with supervised base learner- Decision Tree, classical bagging with unsupervised base learner K-Means, and quantum bagging (QBag) with Q-Means unsupervised base learner approaches under a 5\% label noise setting across the Iris, Breast Cancer, and Wine datasets.
The results highlight that both classical and quantum bagging with unsupervised learner consistently outperform their supervised counterparts in noisy environments. Notably, the Quantum Bagging (QBag) method, which employs Quantum KMeans as its base learner, demonstrates strong robustness and competitive accuracy, particularly in the Breast Cancer and Iris datasets even with minimal reliance on labeled data.

These findings reinforce our central motivation: traditional supervised learners, such as decision trees, are susceptible to label noise and tend to overfit corrupted labels. In contrast, unsupervised base learners are less affected by label noise, as they group data based on its underlying structure rather than relying on potentially corrupted labels.  

\FloatBarrier
\begin{table}[!ht]
    \centering
    \resizebox{\textwidth}{!}
    {%
    \begin{tabular}{|c|c|c|c|c|}
        \hline
        \textbf{Base Learner} & \textbf{No. of Classifiers} & \textbf{Iris (Accuracy)} & \textbf{Breast Cancer (Accuracy)} & \textbf{Wine (Accuracy)} \\
        \hline
        \multirow{8}{*}{Supervised (DT)} 
        & 4  & 0.9133  & 0.8860  & 0.9333  \\
        & 8  & 0.9333  & 0.9175  & 0.9444  \\
        & 12 & 0.9467  & 0.9281  & 0.9556  \\
        & 16 & 0.9333  & 0.9246  & 0.9500  \\
        & 20 & 0.9533  & 0.9246  & 0.9556  \\
        & 24 & 0.9533  & 0.9246  & 0.9556  \\
        & 28 & 0.9667  & 0.9298  & 0.9556  \\
        & 32 & 0.9467  & 0.9316  & 0.9556  \\
        \hline
        \multirow{8}{*}{Classical (KMeans)} 
        & 4  & 0.9400  & 0.9035   & 0.9500  \\
        & 8  & 0.9400  & 0.9333  & 0.9722  \\
        & 12 & 0.9733  & 0.9421  & 0.9722  \\
        & 16 & 0.9667  & 0.9456  & 0.9889  \\
        & 20 & 0.9733  & 0.9439  & 0.9889  \\
        & 24 & 0.9667  & 0.9421  & 0.9889  \\
        & 28 & 0.9733  & 0.9491  & 0.9889  \\
        & 32 & 0.9667  & 0.9421  & 0.9889  \\
        \hline
        \multirow{8}{*}{Quantum (QMeans)} 
        & 4  & 0.9417  & 0.9316  & 0.9167  \\
        & 8  & 0.9750  & 0.9421  & 0.9444  \\
        & 12 & 0.9750  & 0.9474  & 0.9653  \\
        & 16 & 0.9750  & 0.9456  & 0.9792  \\
        & 20 & 0.9667  & 0.9439  & 0.9653  \\
        & 24 & 0.9667  & 0.9456  & 0.9653  \\
        & 28 & 0.9667  & 0.9456  & 0.9792  \\
        & 32 & 0.9583  & 0.9474  & 0.9861  \\
        \hline
    \end{tabular}
    }
    \caption{Comparison of supervised bagging with classical and quantum  bagging using unsupervised base learner under 5\% label noise across three datasets.}
    \label{tab:bagging_comparison}
\end{table}

\section{Conclusion }\label{sec:conclusion}

In this work, we introduced a quantum bagging framework that leverages unsupervised base learners as a noise-resilient alternative to traditional supervised bagging approaches. By integrating QRAM-driven quantum bootstrapping with Delta-K++ initialized quantum clustering, our method bypasses the dependency on labeled data during base learner construction, which is often vulnerable to label noise and inconsistencies in supervised approaches. Unlike conventional bagging methods, which rely on the robustness of the labels, our model instead uncovers structure directly from the data through quantum-enhanced encoding and distance-preserving clustering.
We empirically evaluated our approach across diverse real-world datasets including Breast Cancer, Iris, Wine, and Real Estate Valuation under noisy label and feature conditions. The results demonstrate that the quantum bagging consistently achieves higher generalization performance and lower variance across different numbers of quantum base learners and delta thresholds. This robustness is particularly notable in classification scenarios on the Iris and Wine datasets and regression tasks on the Real Estate dataset, where the quantum bagging maintains stable performance despite the presence of noise.
These findings highlight the potential of unsupervised quantum bagging to offer a scalable, noise-tolerant alternative to supervised bagging learners, especially valuable in domains where clean annotations are limited or unreliable.
\paragraph{Future direction}
Building on the encouraging results under noisy conditions, future work can explore theoretical analysis of variance reduction in quantum unsupervised bagging and extend the framework to more complex or high-dimensional datasets. Extending the method to semi-supervised or active learning scenarios and adapting it for NISQ-era hardware implementations are also compelling directions for expanding the practical impact of quantum bagging.

\bmhead{Acknowledgements}
The first author Neeshu Rathi is thankful to the University Grants Commission(UGC), New Delhi, India, for providing finantial support in the form of a senior research fellowship
with UGC Ref. No.: 1136/(CSIR-UGC NET JUNE 2019).

\section*{Declarations}
\bmhead{Author contributions}
All authors contributed to the study's conception and design.

\bmhead{funding statement}
This work supported by the University Grants Commission(UGC), New Delhi, India, in the form of a senior research fellowship with UGC Ref. No.: 1136/(CSIR-UGC NET JUNE 2019).

\bmhead{Conflict of Interest}
The authors declare no potential conflict of interests.

\bmhead{Ethics approval statement}
This article does not contain any studies with human participants or animals performed by any of the authors.

\bmhead{Consent for publication}
All participants have given their consent for any identifiable information, photos, or data included in this manuscript to be published. Everyone has given permission for these details to be published in this journal. 
\bmhead{permission to reproduce material from other sources}
Not applicable
\bmhead{patient consent statement}
Not applicable
\bmhead{Materials availability}
Not applicable
\bmhead{data availability statement}
Not applicable
\bmhead{clinical trial registration}
Not applicable
\bmhead{Code availability}
The codes are available on request from
the authors.

\bibliography{ref}
\newcommand{\suppmatheader}{
  \newpage
  \begin{center}
    \vspace*{0.5cm}
    \rule{\textwidth}{0.5pt}
    \Huge \textbf{Quantum Bagging}\\[0.2em]
    \Large (Supplementary Material)\\[0.5em]
    \rule{\textwidth}{0.5pt}
    \vspace{0.5cm}
  \end{center}
}

\end{document}